\documentclass[%
 superscriptaddress,
 aip,
 amsmath,amssymb,
reprint,%
 author-year,%
longbibliography
]{revtex4-2}

\usepackage{
amsmath,
amssymb,
bm,
color,
comment,
dcolumn,
enumerate,
float,
graphicx,
hyperref,
mathrsfs,
tabularx,
titlesec,
soul,
subfigure
}

\definecolor{linkColor}{rgb}{0.7,0,0}
\definecolor{darkred}{rgb}{0.7,0,0}
\hypersetup{pdfborder={0 0 0},colorlinks=true,urlcolor=linkColor,citecolor=linkColor}

\titleformat{\section}
{\color{darkred}\sffamily\bfseries}
{\color{darkred}\thesection}{1em}{}
\titleformat{\subsection}
{\color{darkred}\sffamily\itshape}
{\color{darkred}\thesection}{1em}{}
\titlespacing*{\section}
{0pt}{8pt}{0pt}
\titlespacing*{\subsection}
{0pt}{8pt}{0pt}

\emergencystretch=\maxdimen
\begin{document}


\title{Improved Three-Dimensional Reconstructions in Electron Ptychography through Defocus Series Measurements}

\author{Marcel Schloz}
\email{schlmarc@hu-berlin.de}
\affiliation{
Institute of Physics and Center for the Science of Materials Berlin, Humboldt-Universität zu Berlin, Newtonstraße 15, 12489 Berlin,
Germany}

\author{Thomas C. Pekin}
\affiliation{
Institute of Physics and Center for the Science of Materials Berlin, Humboldt-Universität zu Berlin, Newtonstraße 15, 12489 Berlin,
Germany}

\author{Hamish G. Brown}
\affiliation{Ian Holmes Imaging Centre, Bio21 Molecular Science and Biotechnology Institute, University of Melbourne, Melbourne 3010, Australia}

\author{Dana O. Byrne}
\affiliation{Department of Chemistry, University of California, Berkeley, Berkeley, CA 94720, USA}
\affiliation{
National Center for Electron Microscopy, Molecular Foundry, Lawrence Berkeley National Laboratory, Berkeley, CA 94720, USA}

\author{Bryan D. Esser}
\affiliation{Monash Centre for Electron Microscopy, Monash University, Melbourne 3800, Australia}

\author{Emmanuel Terzoudis-Lumsden}
\affiliation{School of Physics and Astronomy, Monash University, Clayton, Melbourne 3800, Australia}

\author{Takashi Taniguchi}
\affiliation{International Center for Materials Nanoarchitectonics, National Institute for Materials Science, 1-1 Namiki, Tsukuba, 305-0044, Japan}

\author{Kenji Watanabe}
\affiliation{Research for Functional Materials, National Institute for Materials Science, 1-1 Namiki, Tsukuba, 305-0044, Japan}

\author{Scott D. Findlay}
\affiliation{School of Physics and Astronomy, Monash University, Clayton, Melbourne 3800, Australia}

\author{Benedikt Haas}
\affiliation{
Institute of Physics and Center for the Science of Materials Berlin, Humboldt-Universität zu Berlin, Newtonstraße 15, 12489 Berlin,
Germany}

\author{Jim Ciston}
\affiliation{
National Center for Electron Microscopy, Molecular Foundry, Lawrence Berkeley National Laboratory, Berkeley, CA 94720, USA}

\author{Christoph T. Koch}
\affiliation{
Institute of Physics and Center for the Science of Materials Berlin, Humboldt-Universität zu Berlin, Newtonstraße 15, 12489 Berlin,
Germany}

\date{\today}
\begin{abstract}

A detailed analysis of ptychography for 3D phase reconstructions of thick specimens is performed. We introduce multi-focus ptychography, which incorporates a 4D-STEM defocus series to enhance the quality of 3D reconstructions along the beam direction through a higher overdetermination ratio. This method is compared with established multi-slice ptychography techniques, such as conventional ptychography, regularized ptychography, and multi-mode ptychography. Additionally, we contrast multi-focus ptychography with an alternative method that uses virtual optical sectioning through a reconstructed scattering matrix ($\mathcal{S}$-matrix), which offers more precise 3D structure information compared to conventional ptychography. Our findings from multiple 3D reconstructions based on simulated and experimental data demonstrate that multi-focus ptychography surpasses other techniques, particularly in accurately reconstructing the surfaces and interface regions of thick specimens.

\end{abstract}
\maketitle

\section*{Introduction}

\begin{figure}
\includegraphics[width=0.99\linewidth]{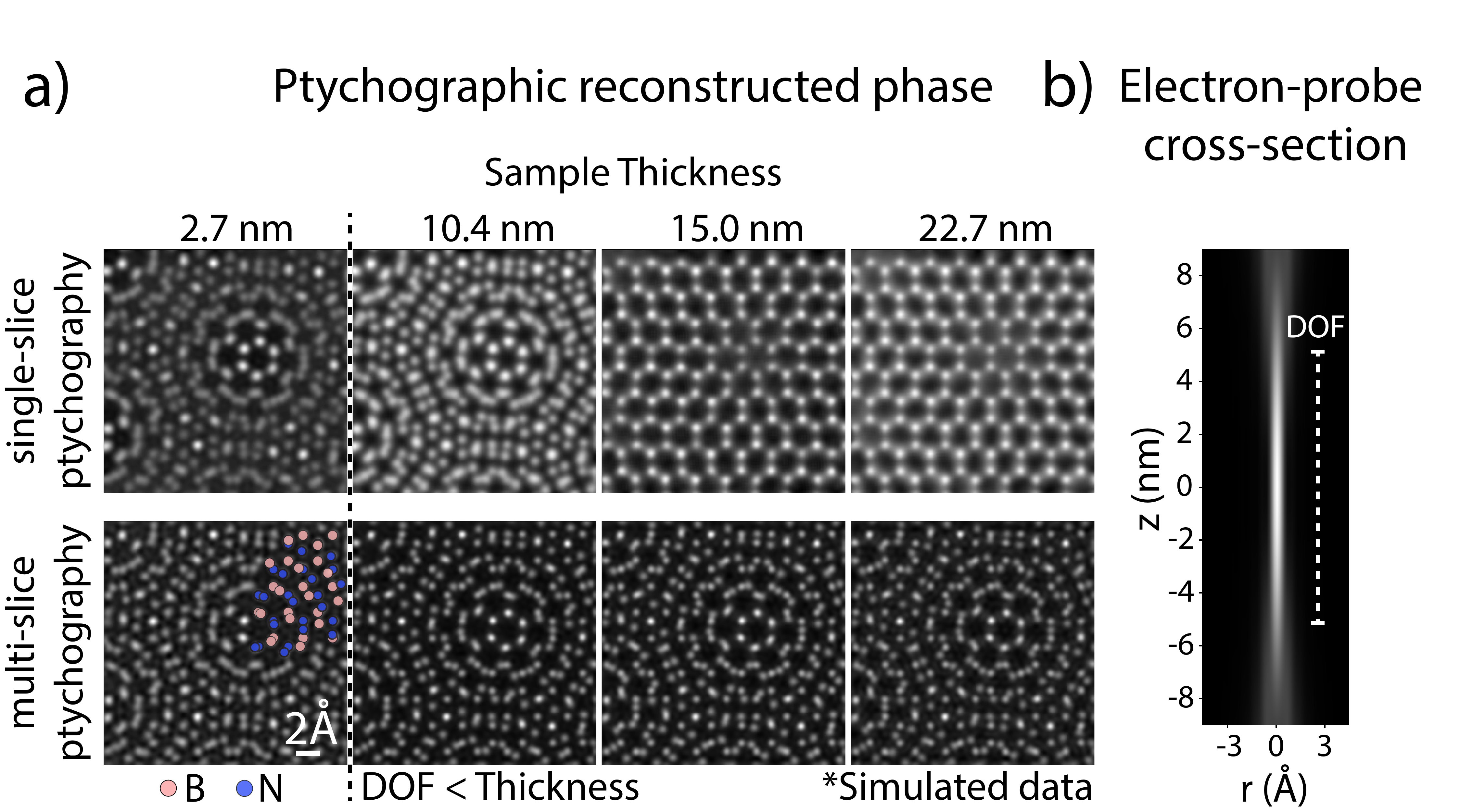}
\centering
\caption[The break-down of the phase object approximation]{The break-down of the phase object approximation. a) Ptychographic phase reconstructions using either a single-slice or a multi-slice model for simulated stacked hBN sample data with different total thicknesses. For the multi-slice model the central slice is shown. The DOF in this setting is about $10.1$~nm. b) Beam profile at $200$~kV and a semi-convergence angle $\theta_{\text{con}}$ of $21$~mrad. This beam is used for the reconstructions in a).}
\label{fig:beam_profiles}
\end{figure}

\begin{figure}
\includegraphics[width=0.99\linewidth]{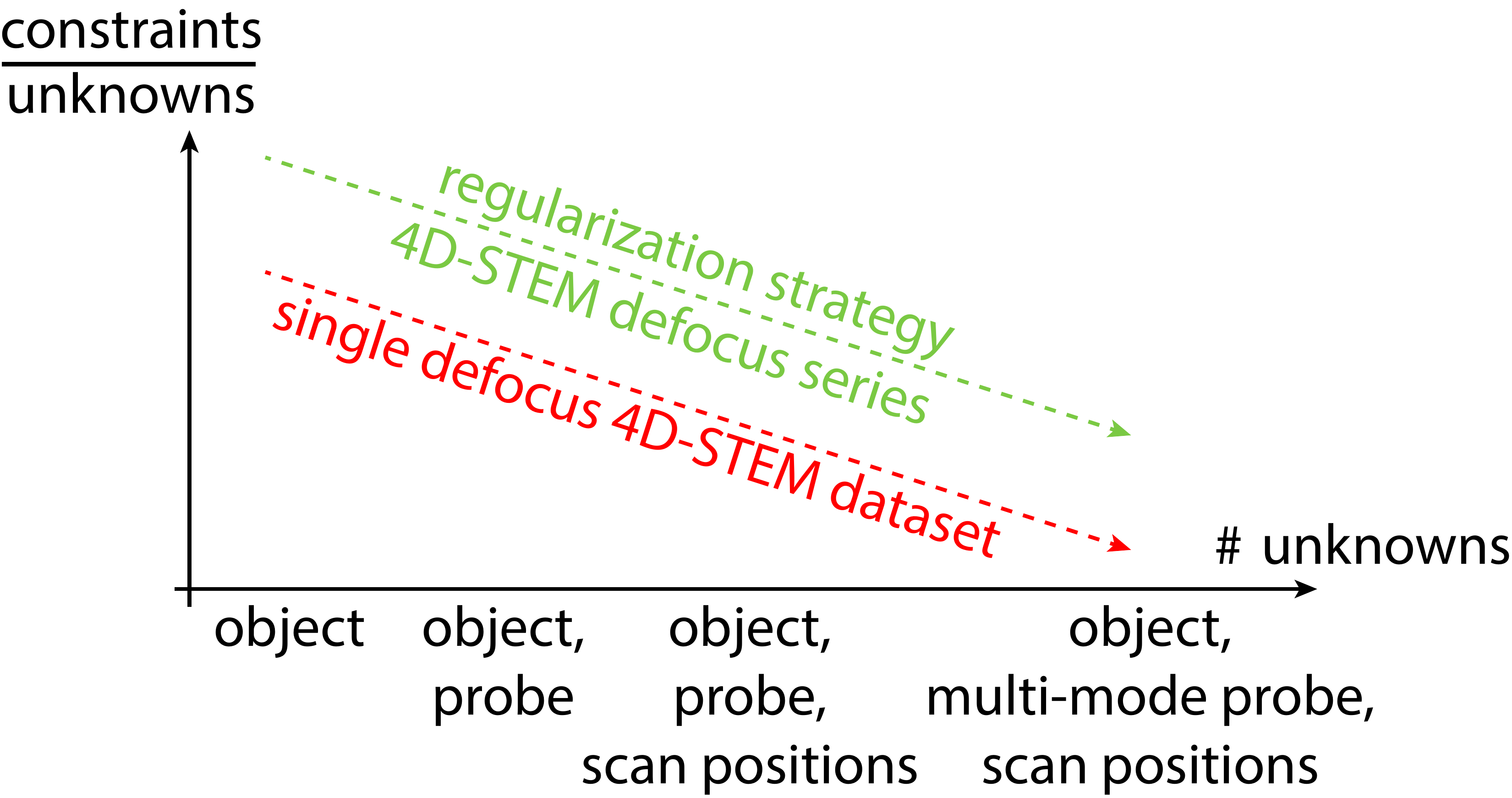}
\centering
\caption[Schematic illustrating the overdetermination
ratio for various multi-slice ptychographic phase retrieval settings.]{Schematic illustrating the overdetermination
ratio for various multi-slice ptychographic phase retrieval settings. Adding more reconstruction tasks, such as adjusting probe, scan positions, or multiple probe modes, can complicate algorithm convergence towards a satisfactory solution. However, incorporating regularization strategies and additional data from defocus series measurements can improve conditioning.}
\label{fig:overdetermination}
\end{figure}

Electron ptychography is a technique whereby the optical phase (itself a function of the electrostatic potential) of the sample is directly reconstructed from a set a diffraction patterns produced by a convergent, coherent electron probe. This technique has achieved dose-efficient quantitative phase imaging at sub-\AA ngstrom resolution, enabling the clear visualization of atomic structures and nanoscale features (\cite{jiang2018electron}).  This capability has significantly contributed to the understanding of various material properties, including crystal defects (\cite{fang2019atomic}), surface reconstructions (\cite{lozano2018low}), and zeolite structures (\cite{zhang2023three}). However, this remarkable resolution is limited to the lateral dimension, constraining reliable structural characterization to 2D-like systems thinner than a few nanometers. Recent research in integrating ptychography into a tilt-series tomography experiment, known as ptychographic atomic electron tomography (PAET), has demonstrated, through simulations (\cite{chang2020ptychographic}) and subsequent experimental validation (\cite{ding2022three, pelz2023solving}), the ability to achieve atomic-resolution phase reconstructions across all three spatial dimensions. However, this gain in depth-resolution comes at the expense of a considerably more challenging experiment and its subsequent analysis. Additionally, the tomographic reconstruction algorithm necessitates specimen projections at a range of tilt angles, provided as single-slice ptychography reconstructions. This constraint again limits the thickness of the investigated specimen to a few nanometers. Extending the PAET technique with multi-slice ptychography holds promise for relaxing the stringent criteria for tomographic sampling (\cite{jacobsen2018relaxation}). This enhancement could potentially simplify experimental acquisition and broaden the technique's applicability to thicker specimens. However, as of now, this capability has not been demonstrated in electron microscopy. \\
\\
The thickness limitation of single-slice ptychography is roughly equivalent to the depth of field (DOF) of the beam (\cite{tsai2016x}). Reconstructions from specimens with thicknesses surpassing the beam's DOF become unreliable due to a breakdown of the phase object approximation, which assumes that the probe and specimen interact in a single, infinitesimally thin plane. For thicker samples, the limited DOF and the increased redistribution of probe intensity caused by scattering make it necessary to use multi-slice ptychography. Figure \ref{fig:beam_profiles}a) shows projected potentials obtained from single-slice and multi-slice reconstructions performed on simulated 4D-STEM data of a two layer stack of hexagonal boron nitride (hBN) specimens. The two layers are twisted with respect to one another, resulting in Moiré patterns at various thicknesses. Additional details about the hBN specimens and the experimental parameters used are provided below. Notably, when the sample thickness exceeds the DOF, single-slice reconstructions show a decrease in resolution and eventually fail to capture the Moiré pattern expected due to the misorientation of the two hBN stacks. In contrast, the resolution remains high in the central slice of the multi-slice reconstruction, and the Moiré patterns are clearly visible. This outcome confirms that the thickness limit does not constrain ptychography when multiple slices are integrated into the model, accounting for multiple scattering effects adequately.
\\
\\
While multi-slice ptychography offers the capability to reconstruct thick specimens, it does come with a drawback: its depth resolution is relatively inferior compared to its lateral resolution. Experimental assessments have demonstrated that achieving a depth resolution slightly better than the aperture-limited depth resolution is feasible (\cite{chen2021electron,o2023three}), indicating a correlation between the depth resolution of multi-slice ptychography and the maximum recorded diffraction angle by the detector (\cite{raines2010three}). However, achieving even this level of depth resolution can be challenging when dealing with experimental 4D-STEM data (\cite{chen2016practical}). In this case, additional experimental parameters, such as the shape and the partial coherence of the probe and the scan positions, often have to be retrieved alongside the object. As a consequence, the number of unknown variables (unknowns) that have to be recovered in the ptychographic iterative phase retrieval process increase for a remaining number of known variables (knowns). This decrease in overdetermination is depicted in the red line in Fig. \ref{fig:overdetermination}. An inadequate ratio of this overdetermination negatively effects the computational stability of the iterative process and leads to incorrect reconstruction results. Measures to alter the ratio of overdetermination favorably are therefore crucial for successfully performing multi-slice ptychography on realistic experimental data. 
\\
\\
Incorporating regularization methods into the reconstruction algorithm has yielded improvements in the conditioning of the reconstruction problem (\cite{thibault2012maximum,schloz2020overcoming,varnavides2023iterative}). One particularly promising approach is the application of a missing-wedge (MW) regularization, which penalizes high axial frequencies at low lateral frequencies in the reconstructed atomic potential in reciprocal space (\cite{chen2021electron}). This regularization thus takes into account the limited influence of the propagation operator in the multi-slice model on the phase of the transmitted electron wave at low lateral frequencies. The effect on the reconstruction result is an elongation of the atoms, a phenomenon also observed in annular dark-field STEM depth-sectioning (\cite{xin2009aberration}) or tilt-series tomography (\cite{midgley2009electron}). While the MW regularization helps the algorithm in addressing the challenges in multi-slice ptychography that arise from the increased number of unknowns in a 3D reconstruction, it is, in principle, less favorable for 3D reconstructions that prioritize high depth-resolution. In an extremely regularized case, the impact of the MW regularization on the reconstruction is akin to a computational approach where all slices are constrained to be identical (\cite{schloz2020overcoming}), resulting in a complete loss of depth information but an enhanced overdetermination ratio.
\\
\\
Another way of improving the ratio of overdetermination is to obtain additional experimental data.
In this paper, we present an extension of multi-slice electron ptychography that involves incorporating a 4D-STEM defocus series. The green line in Fig. \ref{fig:overdetermination} illustrates the ameliorating effect of either applying a regularization strategy or including a 4D-STEM defocus series on the overdetermination ratio in multi-slice ptychography. In contrast to the proposed MW regularization strategy, however, the multi-focus ptychography approach does not negatively impact the achievable depth resolution while keeping the convergence behaviour of the reconstruction algorithm stable. Our implementation of multi-focus ptychography leverages the gradient-based ptychography reconstruction algorithm ROP (\cite{schloz2020overcoming}). Handling the same type of data, our method serves as an alternative reconstruction scheme to the recently introduced $\mathcal{S}$-matrix approach for 3D reconstructions from multi-focus 4D-STEM data (\cite{brown2022three}).

\section*{Materials and methods}
\subsection*{Multi-focus 3D electron ptychography}
Here, we present a novel adaptation of multi-slice ptychography utilizing a set of 4D-STEM datasets acquired at different focal planes within a thick specimen. This experimental approach is akin to conventional STEM depth-sectioning (\cite{voyles2003imaging}), with the distinction that diffraction patterns are recorded at each scan position using a pixelated detector. The various focal lengths in the dataset are chosen to ensure adequate overlap between the DOF of consecutive measurements. By setting a sufficiently large convergence angle to form an electron beam with a small DOF, one can acquire an extensive defocus series, introducing robust redundancy along the axial direction. In the ptychographic reconstruction algorithm, the entire 4D-STEM defocus series is subsequently employed for multi-slice reconstruction. This entails a replacement of the batched loss function, as given in Eq. (15) in Ref. (\cite{schloz2020overcoming}), with:
\begin{equation}
\mathcal{L}(V, \psi^0_0,...,\psi^0_F, \vec{\textit{\textbf{R}}})   =  \frac{1}{P \times F}\sum^F_{f=1}\sum^P_{p=1} \ell(V, \psi^0_f, \textbf{R}_p), \label{eq:batchedLoss_multifocus}
\end{equation}
where the index $f$ corresponds to one of the $F$ defoci used to obtain the series. The gradients of the batched loss given by Eq. (16 - 18) in Ref. (\cite{schloz2020overcoming}) follow accordingly. Note that the gradient of the probe is calculated with respect to each individual defocused probe $\nabla_{\psi^0_f} \mathcal{L}(V,\psi^0_0,...,\psi^0_F,\vec{\boldsymbol{R}})$. The wavefunction $\psi^0_f$ of each 4D-STEM measurement passes through the network model separately and thus the partial derivatives formed by the backward propagation through the model derived in Eq. (12 -  13) in Ref. (\cite{schloz2020overcoming}) independently apply to the gradient of each defocused probe.
\\
\\
In traditional STEM depth-sectioning, a reduced DOF enhances depth resolution by localizing the electron probe, and hence the bulk of the diffraction pattern's information content, to a specific depth range within the specimen. This increased sensitivity to depth variations is particularly useful for detailed investigations of the specimen's structure along the optical axis. Conversely, ptychography does not depend on traditional depth-sectioning to achieve significant depth sensitivity. Since some 3D information is always encoded in the probe wavefunction, in principle, pytchography should be able to reconstruct 3D information about the specimen even from a scan at a single defocus value.
\\
\\
Nevertheless, there are several reasons why multi-focus ptychography could enhance the quality of reconstructions. One contributing factor is the increased number of knowns or rather constraints in this approach, which plays a crucial role in improving reconstruction quality along the optical axis. This becomes particularly significant when paired with enhanced slice sampling in the multi-slice reconstruction, leading to an increase in the number of unknowns. An improved overdetermination ratio could help mitigate any potential ambiguity in the reconstructed atomic 3D potential.\\
\\
Another factor lies in the utilization of 4D-STEM data in multi-focus ptychography, ensuring that all regions of the specimen have been in-focus at least once during the measurement. This aspect is critical as focused-probe ptychography demonstrates greater resilience in scenarios where the illuminating probe is affected by incoherence and/or aberrations. In specimens with a thickness of several tens of nanometers, conventional ptychography leads to out-of-focus regions of the specimen. In the appendix \ref{sec:multi_mode_analysis} of this paper, we demonstrate the necessity of performing reconstructions with multiple probe modes to achieve high reconstruction quality in defocused-probe ptychography. However, optimizing for multiple modes also increases the number of unknowns, making it more challenging for the reconstruction algorithm to generate a reliable solution. This can again be mitigated by the increase in the number of knowns when 4D STEM data from multiple defocus values are available

\subsection*{$\mathcal{S}$-matrix optical sectioning}
This section explores an alternative to multi-focus ptychography for reconstructing 3D structures using a 4D-STEM defocus series: the parallax reconstruction based on the $\mathcal{S}$-matrix technique (\cite{ophus2019advanced, brown2022three, terzoudis2023resolution}). This approach retrieves the atomic potential of thick specimens, which may exhibit heterogeneity along the optical axis, through a two-step process. Initially, the complex elements of the $\mathcal{S}$-matrix are computed from a series of intensity measurements through iterative phase retrieval. Subsequently, the calculated $\mathcal{S}$-matrix is used to generate a virtual through-focal phase series, enabling the determination of a sample's structure in 3D.\\
\\
Suppose we have reconstructed the matrix components $\mathcal{S}_{\textbf{r}_{\perp},\textbf{h}}$, an $\mathcal{S}$-matrix representation transforming an input plane wave into a real-space exit wavefunction. Assuming the projected atomic potential $V^{z_0}(\textbf{r}_{\perp})$ to be in a layer at depth $z_0$ of the specimen, we can describe the scattering process as a free-space propagation to $z_0$, followed by a phase object interaction and further propagation through the remaining distance $\delta z = z_1 - z_0$ gives\footnote{In the special case where the projected potential is located in the mid-plane of the specimen, i.e. $z_0 = z_1/2$, Eq. (\ref{eq:paralax_1}) is identical to a variant of the phase object approximation obtained by using the second-order Strang splitting method (\cite{findlay2021scattering}).} 
\begin{equation}
\label{eq:paralax_1}
\mathcal{S}_{\textbf{r}_{\perp},\textbf{h}} = \mathcal{P}(\textbf{r}_{\perp},\delta z) \otimes_{\textbf{r}_{\perp}}  \left[ e^{i \sigma V^{z_0}(\textbf{r}_{\perp})} e^{-i \pi \lambda h^2 z_0} e^{2 \pi i \textbf{h} \cdot \textbf{r}_{\perp}} \right]. 
\end{equation}
Having obtained the $\mathcal{S}$-matrix through phase retrieval, we can calculate the transmission function at depth $z_0$ by inverting Eq. (\ref{eq:paralax_1}) to:
\begin{equation}
\label{eq:optical_sectioning_result}
 e^{i \sigma V^{z_0}(\textbf{r}_{\perp})}  = \sum_{\textbf{g},\textbf{h} }e^{2 \pi i \textbf{g}\cdot \textbf{r}_{\perp}} e^{i \pi \lambda \delta z g^2 }    \mathcal{S}_{\textbf{g},\textbf{h}} e^{i \pi \lambda  z_0 h^2}  e^{-2 \pi i \textbf{h}\cdot \textbf{r}_{\perp}}.  
\end{equation}
While formulated with the assumption of a sample confined to a single plane $z_0$, the parallax method suggests that the phase component in Eq. (\ref{eq:optical_sectioning_result}) provides an approximation of the structure at various depths $z$ and for various sample thicknesses $\Delta z$. It is important to note that this approximation is not exact. Each slice is treated as a strong phase object and is assumed to affect the measured 4D-STEM intensity independently (\cite{bosch2019analysis}). Despite this assumption of independent contributions from each plane, the parallax reconstruction process at a specific depth, $z$, actually includes elements from multiple planes. However, the influence of elements outside this plane becomes progressively less significant (\cite{brown2022three}). It is also worth noting that errors in determining the scattering matrix (an iterative process) and limitations due to the approximation of Eq. (\ref{eq:optical_sectioning_result}) will both be present in practice and are hard to distinguish.
 
\subsection*{Settings for simulated and experimental data}

\begin{figure}
\includegraphics[width=1.0\linewidth]{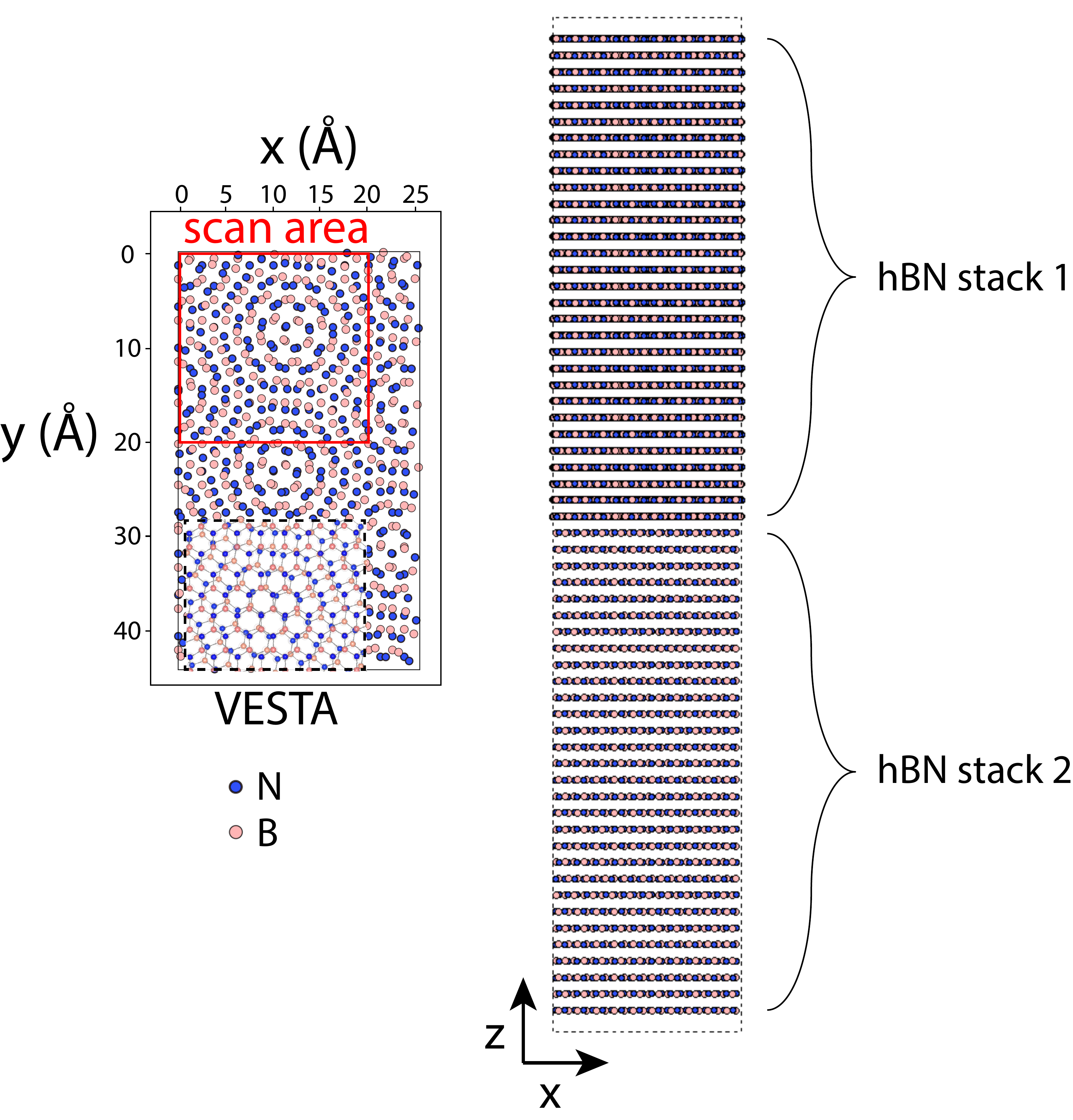}
\centering
\caption[Model of the stacked multi-layer hBN structure.]{Model of the stacked multi-layer hBN structure. The simulated hBN specimen shown from top and side, respectively. The 4D-STEM dataset used for the reconstructions is generated from only a part of this specimen as indicated by the scan area, shown as a red line. To better illustrate the formation of the Moiré pattern in twisted hBN stacks, a VESTA model (\cite{momma2011vesta}) has been added to the top-view representation.
}
\label{fig:sim_hbn_model}
\end{figure}

We explored and compared various multi-slice ptychography approaches (conventional multi-slice ptychography, regularized multi-slice ptychography, and multi-focus ptychography) along with the $\mathcal{S}$-matrix optical sectioning method. We initiated the comparison by applying these multi-slice ptychography methods to an experimental 4D-STEM defocus series dataset obtained previously as described in (\cite{brown2022three}) from a hetero-structure sample comprising lead iridate Pb$_2$Ir$_2$O$_7$ (PIO) and yttrium-stabilized zirconia Y$_{0.095}$Zr$_{0.905}$O$_2$ (YSZ). The estimated thickness of the sample is $200$~\AA, composed of approximately $50$~\AA~of PIO and $150$~\AA~of YSZ. Experimental conditions at the double aberration-corrected
Thermo Fisher Scientific Titan 80-300 microscope involved an accelerating voltage of $300$~kV, a $20$~mrad condenser aperture semi-angle, probe steps of $0.21$~\AA~with a dwell time of $0.874$~ms, and a beam current of $2.01$~pA. Four datasets were acquired using a Gatan K3 direct-electron detector with defocus values $\Delta f$ set to $5.3$~nm, $0$~nm, $-6.2$~nm, and $-14.3$~nm relative to the specimen surface and where we use a positive value of $\Delta f$ to indicate ``overfocus", i.e. a focus point closer to the electron source. The $\mathcal{S}$-matrix approach has previously been applied to reconstruct both light and heavy atoms in this thick sample (\cite{brown2022three}). In this study, alignment of the 4D-STEM datasets was performed using simultaneously acquired HAADF STEM images. The atomic positions of Pb and Ir atoms, as determined from these images, were adjusted to form a distortion-free grid, and this transformation was subsequently applied to the scan positions $\textbf{R}_p$ of the simultaneously-acquired 4D-STEM scan.\\
\\
Besides the PIO-YSZ sample, we conducted both a simulation and an experiment on a Van-der-Waals structure comprising two twisted multi-layered hBN units stacked on top of each other. The simulated structure has a total thickness of approximately $220$~\AA~ and the two stacks were rotated by $10^{\circ}$, while the estimated thickness of the real sample is around $170$~\AA~ and the estimated rotation angle is approximately $12^{\circ}$. For the experimental sample, the twisted structure was created by folding the multilayer hBN unit onto itself, ensuring that each hBN stack has the same thickness. Figure \ref{fig:sim_hbn_model} provides a schematic representation of this structure. This structure serves as an ideal test sample due to its distinct layers of light atoms. However, due to the sensitivity of hBN to knock-on damage (\cite{kotakoski2010electron}), experiments were preferably performed at a limited electron energy. For the experimental investigation, ptychographic datasets were acquired using a NION HERMES microscope (aberration corrected STEM) at $60$~kV (wavelength of $4.87$~pm) acceleration voltage, a $40$~mrad convergence angle $\theta_{\text{con}}$, and a Dectris ELA direct electron detector mounted at the electron energy loss spectroscopy (EELS) camera port. Distortions induced by the EEL spectrometer were corrected using in-house developed software. During data acquisition, a dwell time of $2.0$~ms and a beam current of $19.0$~pA were employed. The defoci of the series ranged from $-12$~nm to $12$~nm with an increment of $4$~nm. For each dataset in the series, a conventional grid scanning procedure with a scanning step size of $0.3$~\AA~was used. Similar to the PIO-YSZ sample, alignment of the 4D-STEM defocus series was performed using simultaneously acquired HAADF-STEM images. However, in this case a method for non-rigid registration of images (\cite{jones2015smart}) was extended to a series of 4D-STEM data sets (\cite{o2022increasing}). The deformations found from synchronously acquired HAADF data were applied to the 4D-STEM data sets by flattening the diffraction space, registering each real-space slice and transforming back to 4D. For the simulation, datasets were generated using \textit{ab}TEM (\cite{abtem}), choosing an acceleration voltage of $60$~kV and a convergence semi-angle $\theta_{\text{con}}$ of $36$~mrad. $106 \times 106$ pixel diffraction patterns were recorded, and a dense grid scan of $100 \times 100$ probe positions with a step size $\Delta x$ of $0.2$~\AA~was chosen. The defocus of the series ranged from $-10$~nm to $10$~nm with an increment of $5$~nm, which sampled the transition between layers sufficiently finely to avoid jumps in the registration due to the change in structure.  For the simulated data used in Fig. \ref{fig:beam_profiles}, the acceleration voltage has been set to $200$~kV and the convergence angle to $21$~mrad, corresponding to a DOF of approximately $10.1$~nm. The diffraction patterns were of size $144 \times 144$ pixel, and a dense grid scan of $67 \times 67$ probe positions with a step size $\Delta x$ of $0.3$~\AA~was used. In all multi-slice reconstructions, 30 slices with a slice distance of $1$~nm were employed. For the simulation of both datasets, a consideration of thermal diffuse scattering (TDS) by using a frozen phonon model in the simulation tool was not included. To ensure comparability with the $\mathcal{S}$-matrix method, the probe positions in the reconstructions were not optimised using the ptychography algorithm.
\section*{Results}
\subsection{Reconstruction from experimental PIO-YSZ data}
\begin{figure}
\includegraphics[width=1.1\linewidth]{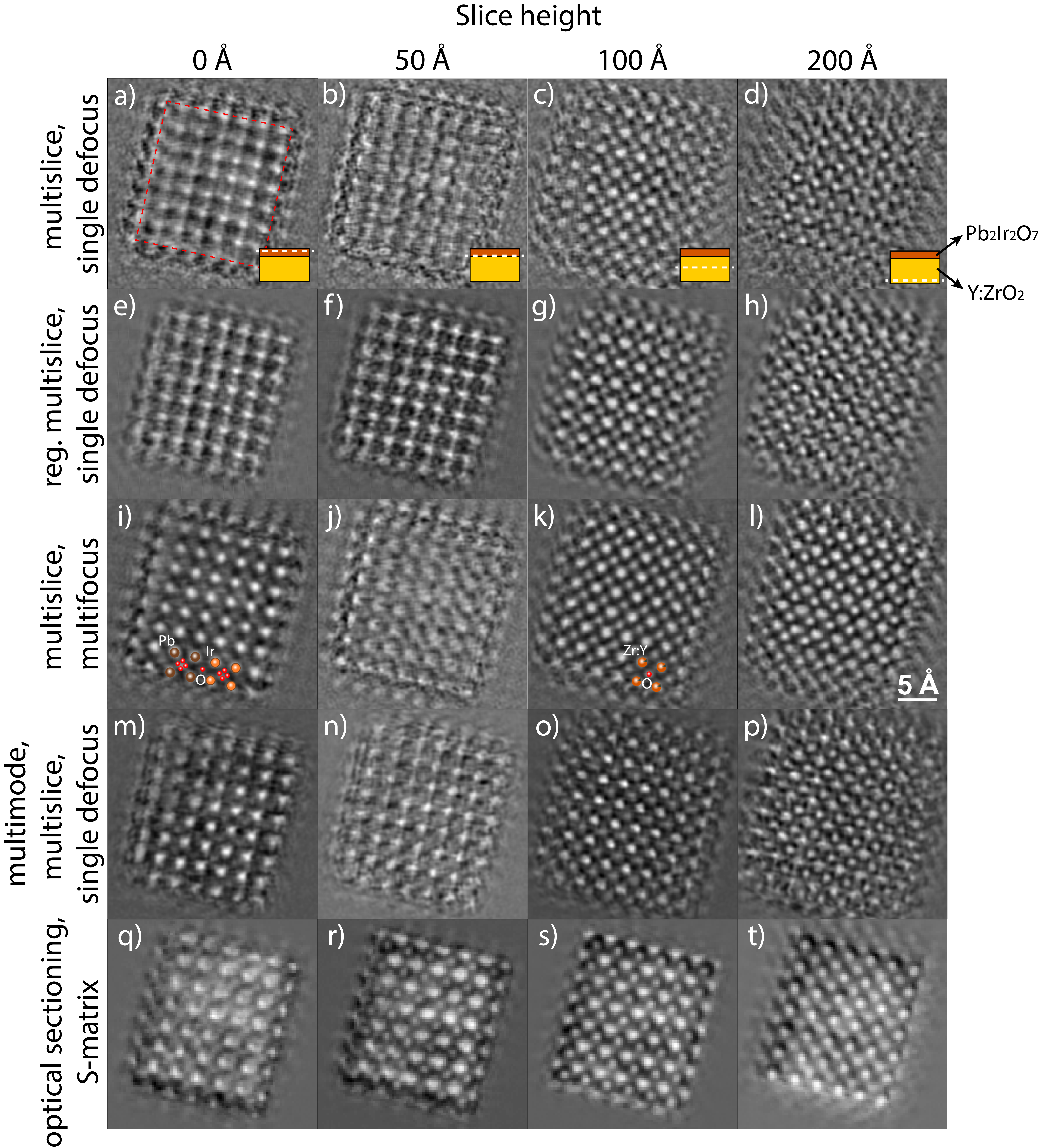}
\centering
    \caption[Ptychographic 3D reconstructions from experimental data of the PIO-YSZ hetero-structure]{Ptychographic 3D reconstructions from experimental data of the PIO-YSZ hetero-structure. a)-d) Reconstructions from multi-slice ptychography using a single dataset with a defocus $\Delta f$ of 5.3~nm. The scanned area is outlined by a red dashed line in a). e)-h) Reconstructions from multi-slice ptychography with MW-regularization applied, using the same dataset as in the non-regularized case. i)-l) Reconstructions from multi-focus, multi-slice ptychography using all four datasets and no MW-regularization. m)-p) Reconstructions from multi-mode, multi-slice ptychography, using again just the same single defocused dataset as in the first two cases. q)-t) Reconstructions obtained from the optical sectioning $\mathcal{S}$-matrix method, also using the entire 4D-STEM defocus series. Adapted with permission from Ref. (\cite{schloz2022high}) \copyright~ Cambridge University Press.}
\label{fig:pio_ysz}
\end{figure}

In this analysis, we conducted a comparative study of the phase reconstructions obtained through conventional multi-slice ptychography, regularized ptychography, multi-focus ptychography, multi-mode ptychography and the $\mathcal{S}$-matrix optical sectioning method. For all ptychography techniques, the atomic potential was reconstructed in ten distinct slices separated by $2.5$~nm in the z-direction. In the conventional multi-slice, regularized and multi-mode approach, only the 4D-STEM dataset acquired at a defocus of $5.3$~nm was utilized. Figure \ref{fig:pio_ysz} displays four out of the ten slices for all ptychography reconstructions. In the same figure four slices generated by the $\mathcal{S}$-matrix optical sectioning approach that correspond to the depth localization of the ptychography slices are presented.

In the conventional ptychography reconstruction, Fig. \ref{fig:pio_ysz}a) clearly shows the termination of the PIO layer, Fig. \ref{fig:pio_ysz}b) indicates the transition between PIO and YSZ with low phase contrast, and Fig. \ref{fig:pio_ysz}c) reveals the uniform crystal structure of YSZ. However, the reduced quality of the reconstructed YSZ crystal lattice in Fig. \ref{fig:pio_ysz}d) suggests challenges in retrieving slices far from the focal plane. Fig \ref{fig:pio_ysz}g) and h) demonstrate an enhancement in the reconstruction quality for YSZ crystal lattice slices through MW-regularization, although at the expense of a less distinct transition between the compounds in Fig. \ref{fig:pio_ysz}f). Multi-focus ptychography, without MW-regularization, overcomes the challenges present in both multi-slice ptychography reconstructions. As shown in Fig. \ref{fig:pio_ysz}i)-l), the high-quality reconstruction reveals detailed structures of each compound and significantly improves the visibility of the transition between the compounds in the hetero-structure. Reconstruction results of multi-mode ptychography are displayed in Fig. \ref{fig:pio_ysz}m)-p). Compared to conventional multi-slice ptychography, this method better resolves the atomic structure of both compounds. However, similar to regularized ptychography, the interface layer shown in n) clearly reveals atomic features, which complicates identifying the transition between the two compounds in the 3D reconstruction. Fig. \ref{fig:pio_ysz}q)-t) displays the reconstruction results obtained with the optical sectioning $\mathcal{S}$-matrix technique. Notably, atoms in all slices appear broader than in the three ptychography results, and some slices exhibit artificial blurring among the atoms. Fig. \ref{fig:pio_ysz}q) representing the crystal structure of the PIO compound is unclear, as it may correspond to the structure of the YSZ compound. However, the appearance of a dip in the reconstructed phase close to the atomic position suggests the presence of high Z atomic columns, i.e. Pb and Ir atoms (Z=82 and Z=77), rather than the lighter Y and Zr atoms (Z=39 and Z=40). Nevertheless, a potential elongation of the YSZ structure to the slice in Fig. \ref{fig:pio_ysz}q) might be possible, given the presence of this structure in the transition slice in Fig. \ref{fig:pio_ysz}r). Thus, while the $\mathcal{S}$-matrix optical sectioning method reveals qualitative differences between the layers, as shown previously (\cite{brown2022three}), the depth resolution is lower than that achieved in multi-slice ptychography.

\begin{figure}
\includegraphics[width=0.99\linewidth]{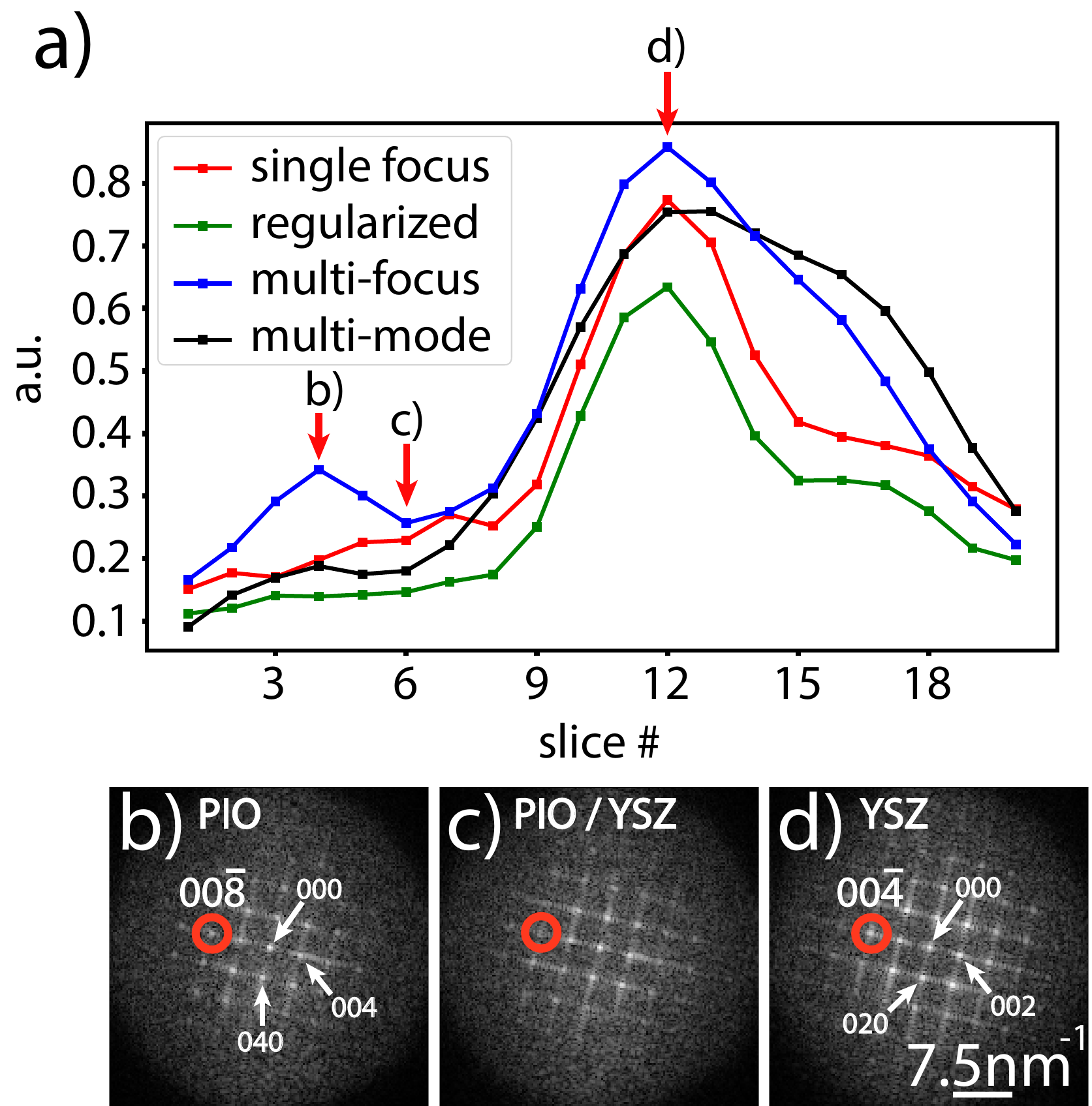}
\centering
    \caption[Depth profile of the PIO-YSZ hetero-structure taken along a diffraction spot in the diffractogram of each 2D slice]{Depth profile of the PIO-YSZ hetero-structure taken along a diffraction spot in the diffractogram of each 2D slice. a) Depth profiles for the four different multi-slice ptychography approaches investigated in this paper (i.e. conventional, regularized, multi-mode and multi-focus ptychography). For multi-focus ptychography, three diffractograms are shown that correspond to b) one of the PIO slices, c) the interface slice and d) one of the YSZ slices. The diffraction spot used to generate the depth profile is encircled in red. }
\label{fig:depth_pio_ysz}
\end{figure}

To further explore the enhancement in quality of ptychographic reconstructions along the beam direction through the inclusion of multiple defocus 4D-STEM datasets, we generated a depth profile at a diffraction spot characteristic of both material compounds, yet absent in the interface layer. Figure \ref{fig:depth_pio_ysz}a) displays the depth profiles for various ptychography techniques, alongside three diffractograms corresponding to the reconstructed PIO, interface, and YSZ slices using the multi-focus technique. Notably, the intensity of the diffraction spot peaks at slice 12, the center of the YSZ compound, across all techniques. However, whereas the depth profiles for conventional, regularized, and multi-mode ptychography show a leveling off towards the PIO compound, the multi-focus technique exhibits a distinct peak, providing a clearer delineation of the PIO compound. For this analysis, the reconstructions were performed using 20 slices, each separated by 1~nm, to enhance the sampling resolution along the optical axis.

\subsection{Reconstruction from simulated and experimental hBN}
\begin{figure}
\includegraphics[width=0.85\linewidth]{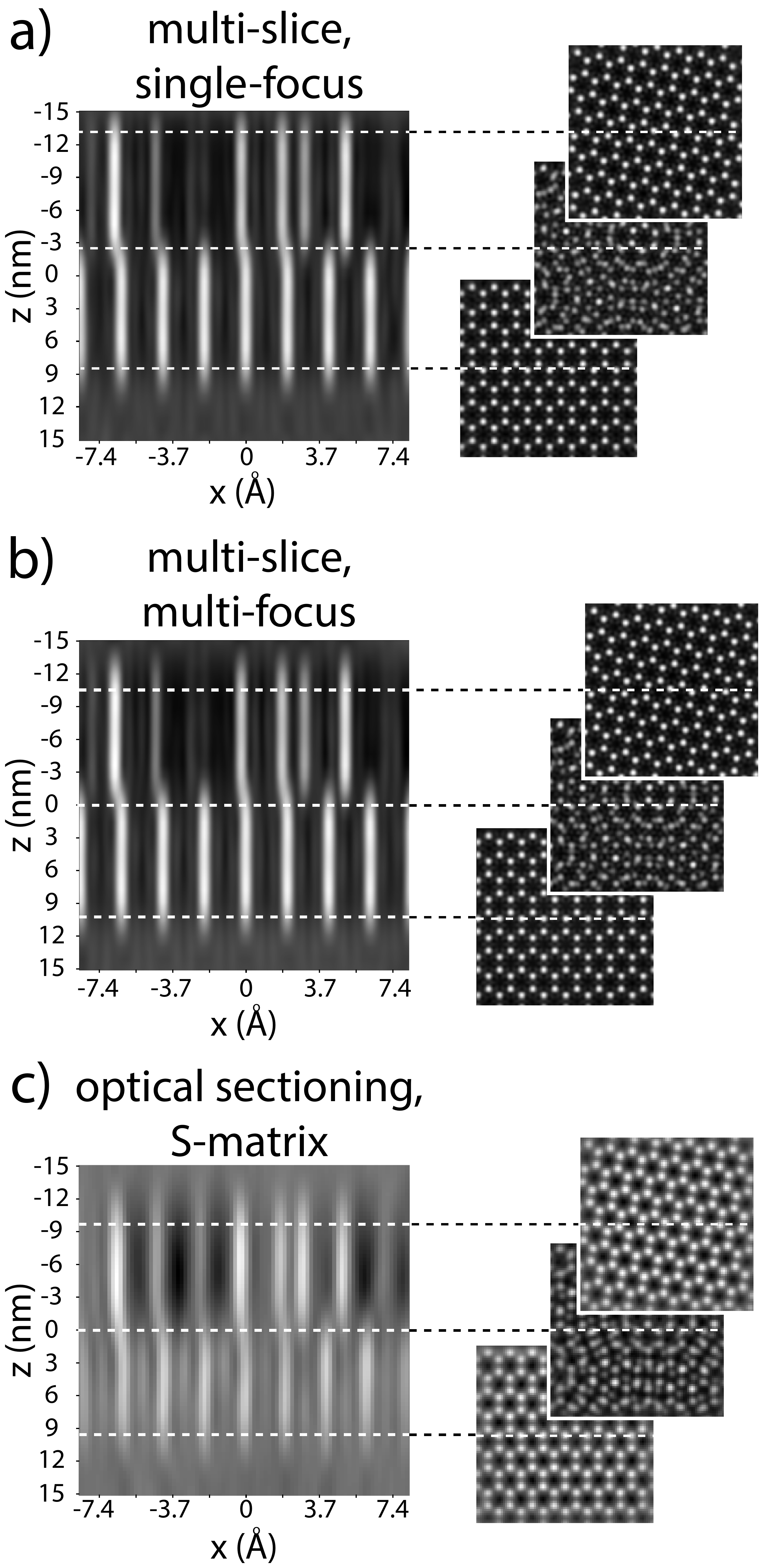}
\centering
\caption[Ptychographic 3D reconstructions from simulated data of a stacked multi-layer hBN structure]{Ptychographic 3D reconstructions from simulated data of a stacked multi-layer hBN structure. A cut-through of a) a  multi-slice ptychography reconstruction using a single 4D-STEM dataset, b) a  multi-slice ptychography reconstruction using a 4D-STEM defocus series and c) the $\mathcal{S}$-matrix optical sectioning reconstruction using the same 4D-STEM defocus series as in b). For each method, three slices are additionally shown that contain only the structure of the first hBN stack, only the structure of the second hBN stack and only the interface of the two stacks.
}
\label{fig:sim_hbn_comparison}
\end{figure}

Figure \ref{fig:sim_hbn_comparison} illustrates the reconstruction of simulated hBN using multi-slice ptychography with a) a single 4D-STEM dataset at a defocus of $0$~nm and b) a 4D-STEM defocus series, along with c) the $\mathcal{S}$-matrix-based optical sectioning that also uses the same defocus series. The 4D-STEM datasets are generated from a scan area depicted in Fig. \ref{fig:sim_hbn_model}, and the three-dimensional reconstructions are presented in Fig. \ref{fig:sim_hbn_comparison}. In all three methods, the two hBN stacks are distinctly separated, and the Moiré pattern resulting from the rotation of the two stacks is visible in the transition slices.

The reconstructions from single-focus and multi-focus ptychography methods exhibit almost identical results, with a slight (3 nm) shift along the optical axis in the former and centering in the latter case. This similarity arises from the full coherence of the beam and the absence of TDS within the sample during the generation of simulated data. With a correct scattering model, there is only one configuration of the potential that can explain the scattered electrons in the data. However, the shift in the reconstruction when a single focus is used indicates that the optimization problem is not adequately constrained. Thus, the voxels corresponding to vacuum can be arbitrarily distributed before or after the reconstructed potential along the optical axis. The fact that the reconstructed specimen is centered along the optical axis in cases where a 4D-STEM defocus series is used supports this assumption.

Comparing the multi-slice ptychography results to the outcome obtained from the $\mathcal{S}$-matrix optical sectioning reveals that the reconstructed atoms are less sharp in the latter case, indicating lower resolution. This effect could stem from the model's inability to handle multiple scattering, a limitation observed in the earlier optical sectioning $\mathcal{S}$-matrix reconstruction of the PIO-YSZ specimen. An alternative interpretation could be based on the fact that ptychography more effectively deconvolves the probe spread function.\\
\\
\noindent The comparison of the three different 3D reconstruction methods is also conducted on the same hBN specimen using experimentally acquired data. 
\begin{figure}
\includegraphics[width=1.0\linewidth]{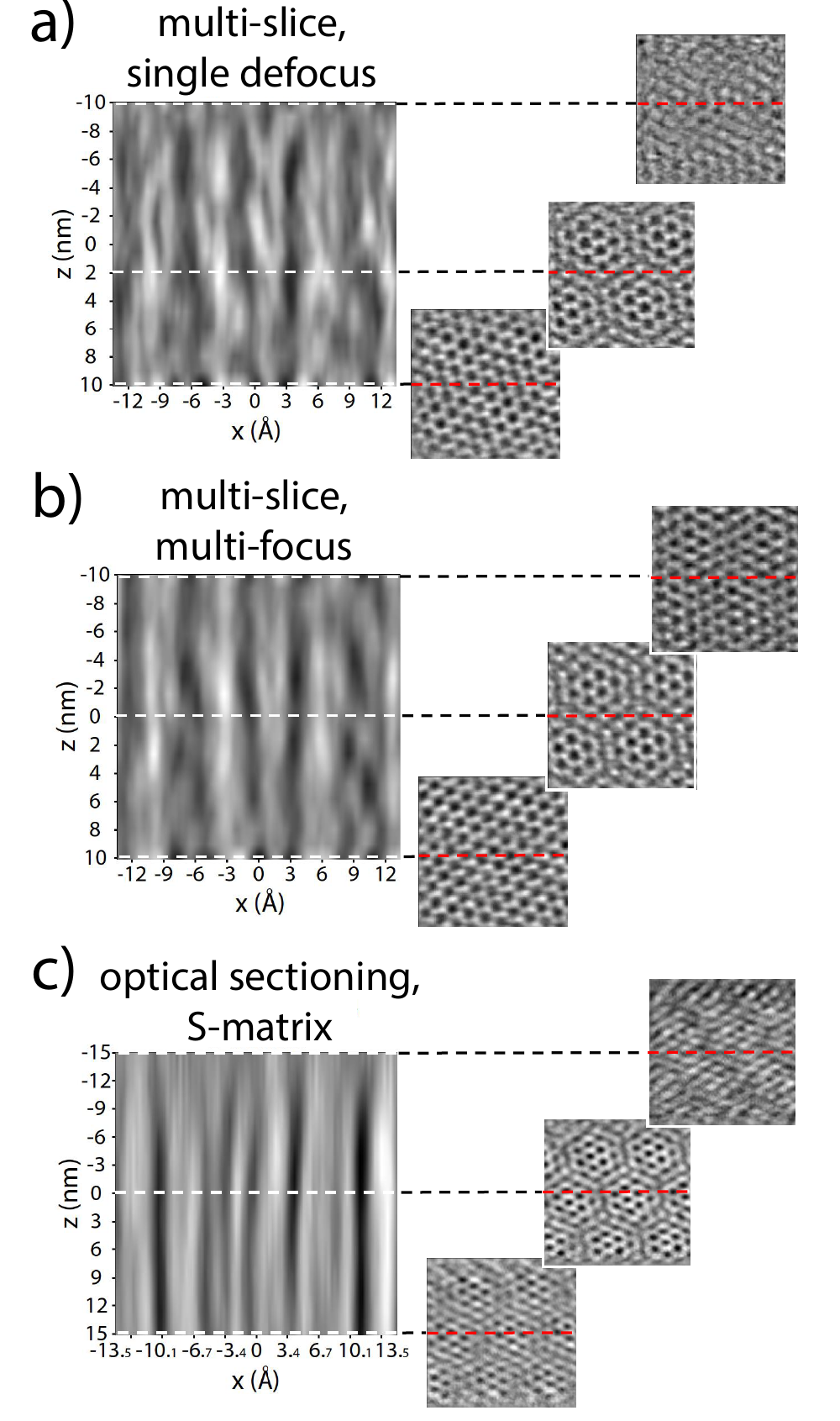}
\centering
\caption[Ptychographic 3D reconstructions from experimental data of a stacked multi-layer hBN structure]{Ptychographic 3D reconstructions from experimental data of a stacked multi-layer hBN structure. A cut-through of the reconstruction generated with a) the conventional ptychographic multi-slice method, b) the ptychographic multi-focus and multi-slice method and c) the $\mathcal{S}$-matrix optical sectioning method. For each method, three slices are additionally shown that best match to the expected structure of the first hBN stack, the expected structure of the second hBN stack and the expected interface of the two stacks.}
\label{fig:exp_hbn_figure}
\end{figure}
Figure \ref{fig:exp_hbn_figure} presents the reconstruction results of the methods. This time, a substantial difference exists between the reconstructions generated by the single-focus and the multi-focus method along the optical axis. In the single-focus reconstruction, the potential fluctuates strongly along the beam direction, whereas this is not the case for the multi-focus method. The increased number of degrees of freedom in the reconstruction algorithm due to the inclusion of multiple slices, combined with realistic experimental conditions, results in incorrect solutions. It becomes evident why MW-regularization is a suitable regularization precisely in this case - high fluctuations along the optical axis are penalized, forcing the reconstruction to closely resemble the one obtained from multi-focus ptychography, where the additional 4D-STEM datasets acquired at different focal planes act as constraints. Similar to Fig. \ref{fig:sim_hbn_comparison}, we observe that the Moiré pattern is not resolved in the central slice of the single-focus reconstruction but rather the $z$-position at which the Moiré occurs is slightly shifted along the optic axis. Moreover, one of the two independent layers of the hBN stack is clearly resolved. In contrast, the structures of each layer in the twisted hBN stack are resolved in the multi-focus reconstruction, and the Moiré pattern is visible in the central slice, where the transition between the two stacks is expected. In the $\mathcal{S}$-matrix optical sectioning result, the transition of the hBN stacks is also clearly visible in the center of the reconstruction. However, the structure of the individual hBN stacks is not resolved since the Moiré pattern dominates the reconstruction along the entire z-direction. This is consistent with a lower axial resolution in the reconstruction which contributes towards an inability to clearly resolve each layer in the twisted structure.

\begin{figure}
\includegraphics[width=0.99\linewidth]{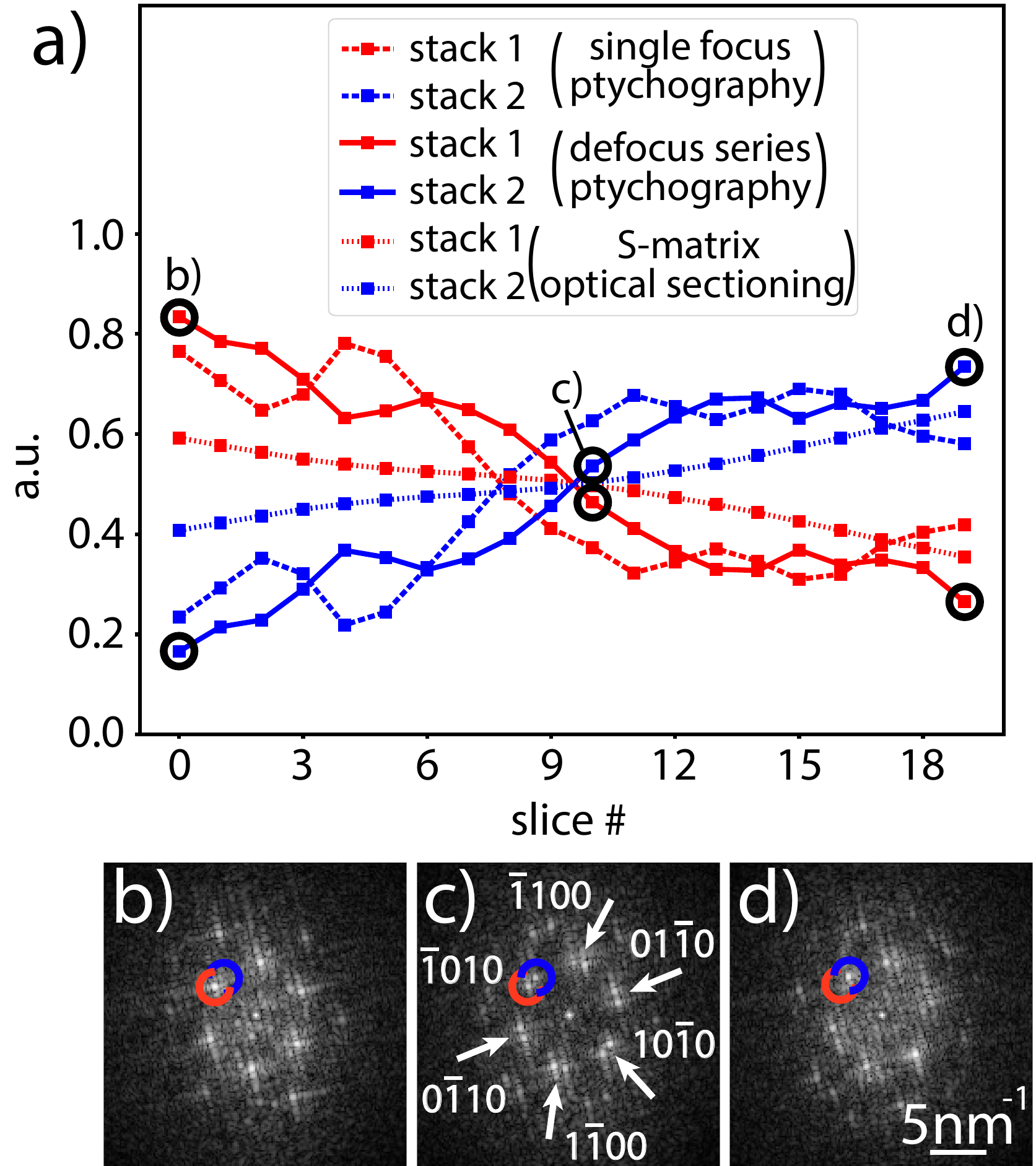}
\centering
    \caption[Depth profiles of the two twisted hBN stacks in the experimental sample, calculated from the average of six diffraction spots in each reconstructed 2D slice's diffractogram]{Depth profiles of the two twisted hBN stacks in the experimental sample, calculated from the average of six diffraction spots in each reconstructed 2D slice's diffractogram. a) shows intensity profiles of the two equally sized hBN stacks using focused-probe ptychography, multi-focus ptychography, and $\mathcal{S}$-matrix optical sectioning. For multi-focus ptychography, b), c), and d) depict diffractograms for specific slices: b) the initial slice of the first hBN stack, c) the interface slice with its Moiré pattern, and d) the final slice of the second hBN stack. A pair of diffraction spots, which contribute to the average used to generate these profiles, is highlighted in red and blue. These colors correspond to the first and second hBN stacks, respectively. The remaining five diffraction spot pairs that form the average are marked with white arrows in c).}
\label{fig:depth_hbn}
\end{figure}

We conducted another analysis on the experimental hBN sample, comparing the intensity ratio of diffraction spots along the beam direction between the two twisted hBN stacks. Figure \ref{fig:depth_hbn}a) presents the averaged intensity of six diffraction spots across three techniques: conventional ptychography using a single focused-probe 4D-STEM dataset, multi-focus ptychography, and $\mathcal{S}$-matrix optical sectioning, both utilizing the 4D-STEM defocus series. Figure \ref{fig:depth_hbn}b)-d) depict diffractograms of the reconstructed slices from multi-focus ptychography, specifically at the onset of the first hBN stack, the interface, and the onset of the second hBN stack. These images illustrate the contributions from each stack. The depth profile from conventional ptychography reveals a relatively uneven intensity trend across the diffraction spot pairs. As previously noted in Figure \ref{fig:exp_hbn_figure}, the interface layer, characterized by its distinct Moiré pattern, is not perfectly centered within the reconstructed volume. In contrast, the depth profile from the multi-focus ptychography demonstrates a smoother transition between the two stacks, with a significantly higher ratio at the beginning and end of the reconstruction, indicating a more accurately reconstructed separation. Meanwhile, the depth profile from the $\mathcal{S}$-matrix optical sectioning reconstruction shows a highly linear trend. However, the ratio towards the surfaces of the reconstructed sample remains relatively low, making it less effective in pinpointing the exact locations of the hBN stacks within the volume.

The 3D reconstruction results obtained with the multi-focus and multi-slice ptychography method and the $\mathcal{S}$-matrix optical sectioning method from the experimental data could potentially be enhanced through an alternative postprocessing of the 4D-STEM defocus series. The currently used HAADF-STEM-based approach for dataset alignment is sub-optimal because HAADF images are low in contrast when the specimen is thin and primarily composed of atoms with a low atomic number, especially when the focal plane does not lie within the specimen. However, restricting the defocus range to only fall within the specimen thickness would make it more challenging for the reconstruction algorithm to correctly recover the surface boundaries. Therefore, we suggest that an alternative alignment of the defocus series through a multi-slice ptychography reconstruction of each individual dataset may potentially help address the limitations of the current alignment procedure.

\section*{Conclusion}
An extension of multi-slice ptychography has been proposed, leveraging a 4D-STEM defocus series from a single sample orientation to improve the three-dimensional phase reconstruction. The method was validated through simulations and experiments involving various samples, each with a thickness ranging up to tens of nanometers and atomic numbers ranging from 5 (B) to 82 (Pb). More specifically, the method was applied to experimental data involving a $20$~nm thick PIO-YSZ hetero-structure and simulated and experimental data for a $22$~nm and $17$~nm thick multi-layered hBN sample, respectively. Incorporating of a defocus series into the ptychographic reconstruction algorithm mitigates the impact of partial incoherence on the reconstruction outcome, rendering the use of regularization methods unnecessary for achieving a correct solution. This innovation has demonstrated higher fidelity of reconstructions along the beam direction as evidenced by the smoothness and overall value of reconstructed potential in homogeneous areas. The capacity to perform three-dimensional phase reconstructions using a 4D-STEM defocus series is also achievable with an $\mathcal{S}$-matrix-based optical sectioning approach. We conducted a comparative analysis of the two approaches, revealing that the implicit treatment of multiple scattering through averaging over different momentum components, as employed in the $\mathcal{S}$-matrix approach, is inadequate. Instead, an explicit inclusion of multiple scattering in the model, characteristic of multi-slice ptychography, proves essential for a meaningful reconstruction. 

\appendix
\section{Analysis of multi-mode ptychography}
\label{sec:multi_mode_analysis}
\begin{figure*}
    \centering
    \includegraphics[width=0.99\textwidth]{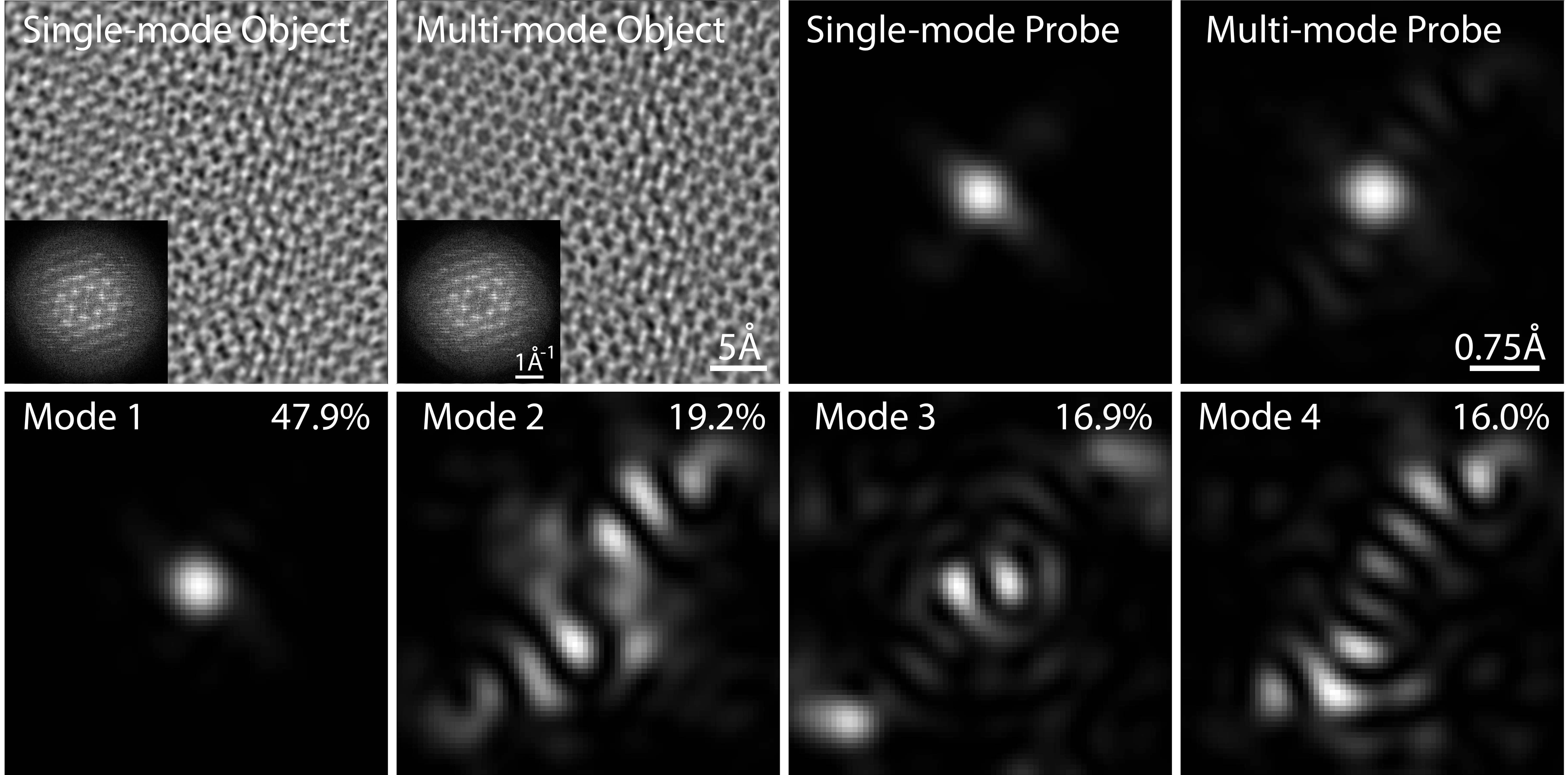}
    \caption[Comparison between the phase reconstructions of a stacked multilayer hBN sample obtained from single-mode and multi-mode ptychography]{Comparison between the phase reconstructions of a stacked multilayer hBN sample obtained from single-mode and multi-mode ptychography. The central slice of the multi-slice reconstructions and their respective diffractograms are shown. The focused probes used in both reconstructions are illustrated and the four probe modes that form the multi-mode probe with their respective contribution are given.}\label{fig:multimode_experimental_hbn}
\end{figure*}

\begin{figure*}
    \centering
    \includegraphics[width=0.99\textwidth]{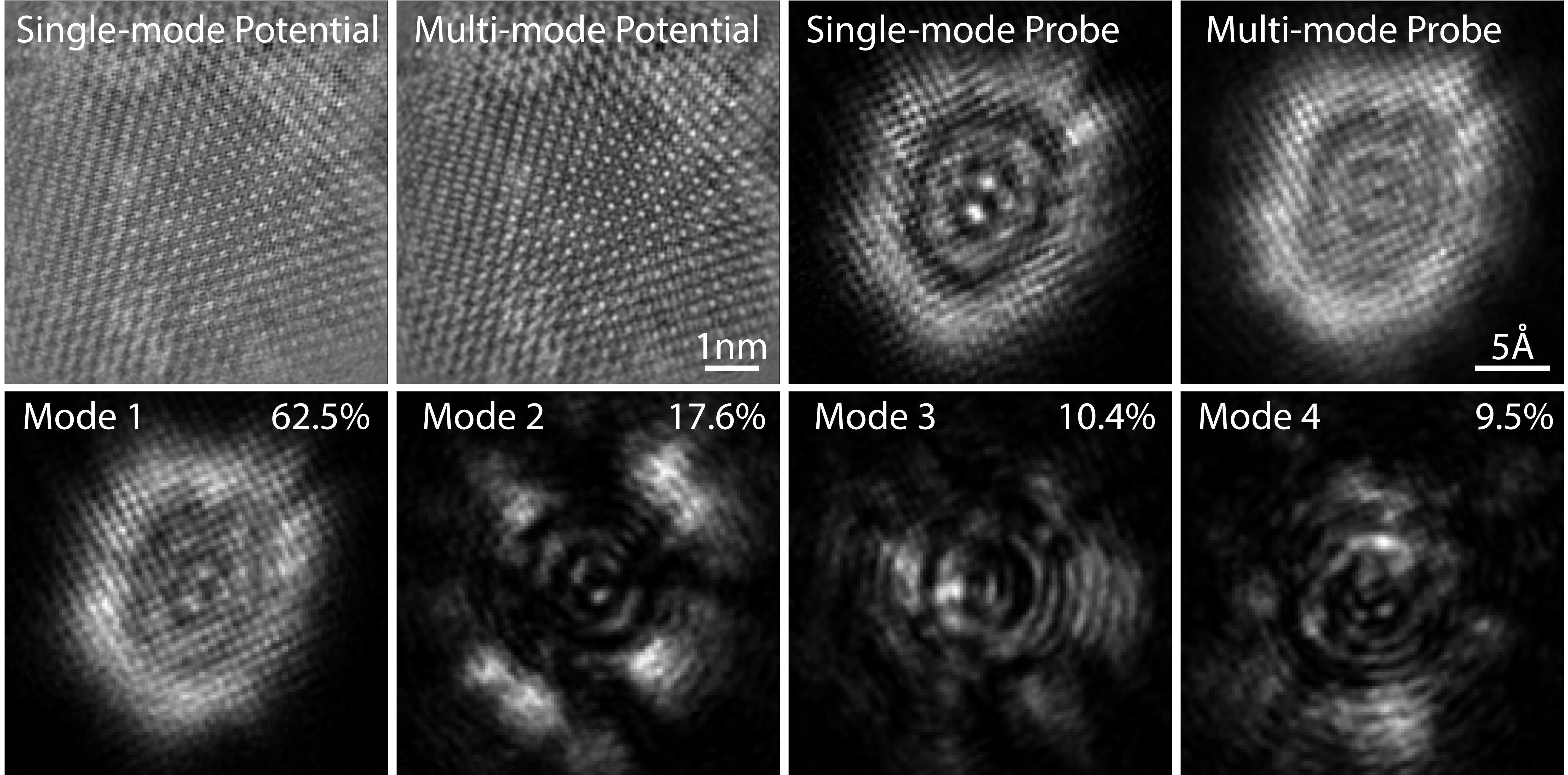}
    \caption[Improvement of the phase reconstruction of a bilayer MoSe$_2$/WS$_2$ sample through multi-mode ptychography]{Improvement of the phase reconstruction of a bilayer MoSe$_2$/WS$_2$ sample through multi-mode ptychography. The reconstructions are performed with single-mode and multi-mode ptychography and the corresponding defocused probes of the two reconstructions are illustrated. The four probe modes with their respective contribution to the multi-mode probe are shown.}\label{fig:multimode_experimental_zhen}
\end{figure*}

In this study, we investigate the impact of partial spatial coherence in the illuminating probe on ptychographic reconstructions when using just a single 4D-STEM dataset and explore the extent to which a mixed-state model can counteract it. To achieve this, we conducted a comparative analysis between a single-mode and a multi-mode ptychographic reconstruction using an experimentally acquired dataset of the stacked multi-layered hBN sample, previously detailed in this paper. In this case, the experiment was conducted with a Thermo Fisher Scientific Spectra $\phi$ FEGTEM microscope fitted with a CEOS SCORR+ probe spherical aberration corrector, operating at $300$~kV (wavelength of $1.97$~pm) acceleration voltage, a semi-convergence angle $\theta_{\text{con}}$ of $25$~mrad, and an EMPAD direct electron detector. The selected sample region exhibited a Moiré-like pattern arising from a slight rotation of the two hBN stacks, covering the entire field of view (FOV) of $3.33$~nm $\times$ $3.33$~nm, corresponding to $256 \times 256$ scan positions. Data acquisition involved a dwell time of $1.0$~ms and a beam current of $6.1$~pA. In the multi-mode reconstruction, we employed four different Hermite-Gaussian probe modes, orthogonalized using the Gram-Schmidt orthogonalization method. A multi-slice model with five slices and a slice distance of $3.6$~nm was utilized to accommodate the estimated sample thickness of approximately $17$~nm. The reconstruction algorithm optimized the object and probe modes over 400 iterations. Figure \ref{fig:multimode_experimental_hbn} displays the reconstructed slice of the interface between the two hBN stacks and the illuminating probe for both the single-state and mixed-state models.
The retrieved probes are depicted for both scenarios, alongside the four individual probe modes and their respective contributions to the multi-mode probe, expressed as a percentage. A comparison between single-mode ptychography and multi-mode ptychography reveals an enhancement in reconstruction quality with the latter. Specifically, the hBN structure, particularly the unit cells, is more clearly resolved. However, when examining their diffractograms, the resolutions of the two reconstructions appear generally similar. Consequently, while the present findings show some advantage of multi-mode ptychography over convetional single-mode ptychography, the substantial difference in reconstructions reported in Ref. (\cite{chen2020mixed}) and (\cite{chen2021electron}) was not replicated in this study.\\
\\
To further investigate, we conducted an additional analysis using the bilayer MoSe$_2$/WS$_2$ data from Ref. (\cite{chen2020mixed}). Once again, we performed both single-mode and multi-mode ptychographic reconstructions, employing four orthogonalized probe modes for the latter. Figure \ref{fig:multimode_experimental_zhen} displays the single-slice reconstruction of the sample alongside the single-mode probe used, and the single-slice reconstruction with the multi-mode probe and its probe modes. In this comparison, the difference between the reconstruction results is quite pronounced, consistent with the reported results. Structural features, such as a monolayer of WS$_2$ and well-aligned and misaligned stacking in bilayer MoSe$_2$/WS$_2$ regions, are much more clearly resolved with multi-mode ptychography. Notably, the experiment involved a strongly defocused probe of approximately $55$~nm and a large scan step size of $2.36$~\AA~, in contrast to the focused-probe experiment in the previous analysis. This implies that partial spatial incoherence has a more pronounced impact on reconstruction quality when using a defocused probe. This finding is in agreement with what has been reported in Ref. (\cite{chen2020mixed}).

\section*{Acknowledgments}
M.S., T.C.P., and C.T.K acknowledge support from the Deutsche Forschungsgemeinschaft
(DFG, German Research Foundation) - Project-ID 414984028 - SFB 1404. Work at the Molecular Foundry was supported by the Office of Science, Office of Basic Energy Sciences, of the U.S. Department of Energy under Contract No. DE-AC02-05CH11231. D.O.B. also acknowledges funding from the Department of Defense through the National Defense Science \& Engineering Graduate (NDSEG) Fellowship Program. J.C. acknowledges additional support from the Presidential Early Career Award for Scientists and Engineers (PECASE) through the U.S. Department of Energy. This research was supported under the Discovery Projects funding scheme of the Australian Research Council (project no. FT190100619). This research was supported by an Australian Government Research Training Program Scholarship.

\section*{References}

\bibliography{refs}    

\end{document}